\documentclass[aps,prl,reprint,twocolumn,superscriptaddress,amsmath,amssymb]{revtex4-2}
\usepackage{graphicx,bm,mathptmx,physics,color}
\usepackage[colorlinks=true,allcolors=blue]{hyperref}
\usepackage[sort&compress]{natbib}

\begin{document}
\title{Instantaneous Quasiphoton in Strong-Field Ionization}

\affiliation{State Key Laboratory of Precision Spectroscopy, East China Normal University, Shanghai 200241, China}
\affiliation{Key Laboratory for Laser Plasmas (Ministry of Education) and School of Physics and Astronomy, Collaborative Innovation Center for IFSA (CICIFSA), Shanghai Jiao Tong University, Shanghai 200240, China}
\affiliation{Collaborative Innovation Center of Extreme Optics, Shanxi University, Taiyuan, Shanxi 030006, China}
\affiliation{CAS Center for Excellence in Ultra-intense Laser Science, Shanghai 201800, China}

\author{Yongzhe~Ma}
\affiliation{State Key Laboratory of Precision Spectroscopy, East China Normal University, Shanghai 200241, China}
\author{Hongcheng~Ni}
\email{hcni@lps.ecnu.edu.cn}
\affiliation{State Key Laboratory of Precision Spectroscopy, East China Normal University, Shanghai 200241, China}
\affiliation{Collaborative Innovation Center of Extreme Optics, Shanxi University, Taiyuan, Shanxi 030006, China}
\author{Yang~Li}
\affiliation{Key Laboratory for Laser Plasmas (Ministry of Education) and School of Physics and Astronomy, Collaborative Innovation Center for IFSA (CICIFSA), Shanghai Jiao Tong University, Shanghai 200240, China}
\author{Feng~He}
\affiliation{Key Laboratory for Laser Plasmas (Ministry of Education) and School of Physics and Astronomy, Collaborative Innovation Center for IFSA (CICIFSA), Shanghai Jiao Tong University, Shanghai 200240, China}
\affiliation{CAS Center for Excellence in Ultra-intense Laser Science, Shanghai 201800, China}
\author{Jian~Wu}
\email{jwu@phy.ecnu.edu.cn}
\affiliation{State Key Laboratory of Precision Spectroscopy, East China Normal University, Shanghai 200241, China}
\affiliation{Collaborative Innovation Center of Extreme Optics, Shanxi University, Taiyuan, Shanxi 030006, China}
\affiliation{CAS Center for Excellence in Ultra-intense Laser Science, Shanghai 201800, China}

\begin{abstract}
Photon is a concept that does not apply at the instantaneous level when light is described by classical electromagnetic fields. Exploiting the dynamical rotational symmetry of circularly or elliptically polarized classical light pulses, however, we demonstrate the existence of instantaneous quasiphotons down to the subcycle level. We illustrate the concept of quasiphotons in strong-field ionization through the correlated spectrum of angular momentum and energy of photoelectrons, both at the tunnel exit and in the asymptotic region. Moreover, we propose a protocol based on electron vortices to directly visualize the existence of quasiphotons. Our work paves the pathway towards a deeper understanding of fundamental light-matter interactions with photonic characteristics on the subcycle scale.
\end{abstract}

\maketitle

In strong-field ionization, where the number of photons associated with the quantum state of light is large, the state remains largely unperturbed upon light-matter interaction \cite{Glauber1963}. Thereby, light can usually be described by classical optical fields in strong-field ionization. Nevertheless, distinct photonic characteristics emerge from the optical fields. A notable example is the above-threshold ionization (ATI) \cite{Agostini1979,Eberly1991,Becker2002,Milosevic2006}, where the photoelectron energy spectrum exhibits prominent discrete peaks separated by one photon energy. The interpretation of photonic ATI peaks arising from classical optical fields can be formalized within the framework of Floquet theory \cite{Burnett1993,Joachain2012}. For a time-periodic Hamiltonian, $H(t+T)=H(t)$, with a period $T =2\pi/\omega$, Floquet quasienergy states emerge, exhibiting an energy separation of $\omega$. The ATI peaks can be viewed as the manifestation of these Floquet states \cite{Chu1985,Potvliege1988,Chu2003,Smirnova2008}, thereby introducing the concept of photons in the presence of classical optical fields.

When described classically, light can be recognized either as photons in the frequency domain, or as classical electromagnetic waves in the time domain. The uncertainty principle across the time-frequency phase space dictates that the photon is a concept that does not apply at one particular time instance, or, at the instantaneous level. This very fact is often reflected in the well-known uncertainty relation between particle number and phase \cite{Carruthers1968}. Recently, it has been explicitly shown that a minimal of two optical cycles is needed for the Floquet states to emerge \cite{Lucchini2022}, ruling out the possibility to define photons down to the instantaneous level.

\begin{figure}[b]
\centering
\includegraphics[width=\columnwidth]{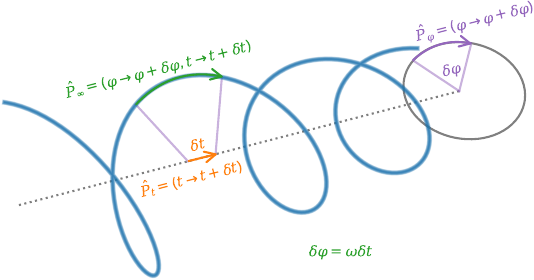}
\caption{Illustration of the dynamical symmetry in a circularly polarized laser field. The Hamiltonian is invariant under an arbitrary time translation $\hat{P}_t=(t\rightarrow t+\delta t)$ combined with a rotation operation $\hat{P}_{\varphi}=(\varphi\rightarrow\varphi+\delta\varphi)$ with $\delta\varphi=\omega\delta t$. Consequently, an infinite-order dynamical symmetry $\hat{P}_\infty=\hat{P}_{\varphi}\hat{P}_t$ emerges, providing support for the introduction of instantaneous quasiphotons on the subcycle scale.}
\label{fig:QuasiFloquet}
\end{figure}

When a system possesses dynamical symmetries, the Floquet theory facilitates discoveries of additional conservation laws and selection rules \cite{Alon1998,Neufeld2019}. For instance, when atoms interact with linearly polarized light, the Hamiltonian exhibits a second-order dynamical symmetry, i.e., it is invariant under the symmetry operation $\hat{P}_2=\hat{P}_{-\bm{r}}\hat{P}_{T/2}$, where $\hat{P}_{-\bm{r}}=(\bm{r}\rightarrow-\bm{r})$ and $\hat{P}_{T/2}=(t\rightarrow t+T/2)$, involving a half-period time translation together with a space inversion operation. This second-order dynamical symmetry has been demonstrated to lead to odd-order harmonics in high harmonic generation \cite{Alon1998}. Remarkably, when atoms interact with circularly polarized light, the Hamiltonian possesses an infinite-order dynamical symmetry, as depicted in Fig.~\ref{fig:QuasiFloquet}. In other words, it remains invariant under an arbitrary time translation $\hat{P}_t=(t\rightarrow t+\delta t)$ accompanied by a rotation operator with a corresponding angle $\hat{P}_{\varphi}=(\varphi\rightarrow\varphi+\delta\varphi)$:
\begin{equation}
\hat{P}_\infty=\hat{P}_{\varphi}\hat{P}_t=(\varphi\rightarrow\varphi+\delta\varphi,t\rightarrow t+\delta t). \quad \delta\varphi=\omega\delta t
\label{eq:DS_Pinf}
\end{equation}
This infinite-order dynamical symmetry points to the possibility of defining quasiphotons that hold validity even at any time instant within an optical cycle.

In this Letter, we illustrate the concept of instantaneous quasiphotons on the subcycle scale through the correlated spectrum of angular momentum and energy (SAME) of photoelectrons, both at the tunnel exit and in the asymptotic region, in strong-field ionization of atoms using circularly and elliptically polarized light pulses. Angular momentum and energy are chosen for the correlated spectrum as they correspond to the conjugate variables of rotation and time translation operations, respectively. We demonstrate that quasiphotons give rise to prominent conservation laws between the angular momentum and energy of photoelectrons. Additionally, we propose a protocol based on the interference-induced electron vortices for direct visualization of the concept of quasiphotons.

To construct the SAME of photoelectrons at the tunnel exit, we perform numerical simulations using the time-dependent Schr\"odinger equation (TDSE) for the helium atom within the single-active electron approximation \cite{Tong2005}, interacting with an intense laser pulse described by the classical vector potential defined in the $x$-$y$ polarization plane as
\begin{equation}
\bm{A}(t)=A_0 \cos^4\left(\frac{\omega t}{2N}\right)
\begin{pmatrix}
\cos(\omega t)\\
\varepsilon \sin(\omega t)
\end{pmatrix},
\label{eq:laser}
\end{equation}
where $A_0$ is the magnitude of the vector potential in the major $x$ axis of the polarization plane corresponding to a laser intensity of $8\times10^{14}$ W/cm$^2$, $\omega$ is the angular frequency corresponding to a wavelength of 800 nm, $N=20$ is the total number of optical cycles, and $\varepsilon=1$ or $0.7$ is the ellipticity. The corresponding laser electric field is defined as $\bm{F}(t)=-\dot{\bm{A}}(t)$. For the given laser parameters, the Keldysh parameter is 0.72 for the circular case and 0.62 for the elliptical case, indicating ionization in the tunneling regime. To solve the TDSE, we employ the split-operator Fourier method on a grid with 2048 points in each dimension, a grid step of 0.2 a.u., and a time step of 0.02 a.u. To suppress reflection of the outgoing wave function, an absorbing boundary of the form $1/[1+\exp\{(r-r_0)/d\}]$ is placed near the grid border, where $r_0=189.8$ and $d=4$ a.u. In order to retrieve information about the tunneling dynamics at the tunnel exit, we employ the backpropagation method \cite{Ni2016,Ni2018,Ni2018a,Hofmann2021,Ma2023}. This hybrid quantum-classical method treats the tunneling dynamics fully quantum mechanically and retrieves the tunneling exit characteristics from the eventually ionized portion of the wave function through backpropagation along the time axis using classical electron trajectories. The backpropagation method has been extensively applied in various studies, including the retrieval of tunneling exit characteristics \cite{Klaiber2022}, extraction of tunneling \cite{Wang2017,Zhang2017} and deformation dynamics \cite{Liu2018} of atomic $p$ orbitals, probe of rescattering time \cite{Kim2021}, and investigation of subcycle transfer of linear momentum \cite{Ni2020} due to nondipole effects \cite{Smeenk2011,Wang2020,Maurer2021,Mao2022}. It has been shown to be a reliable approach for obtaining highly differential information about the tunnel exit.

\begin{figure}[t]
\centering
\includegraphics[width=\columnwidth]{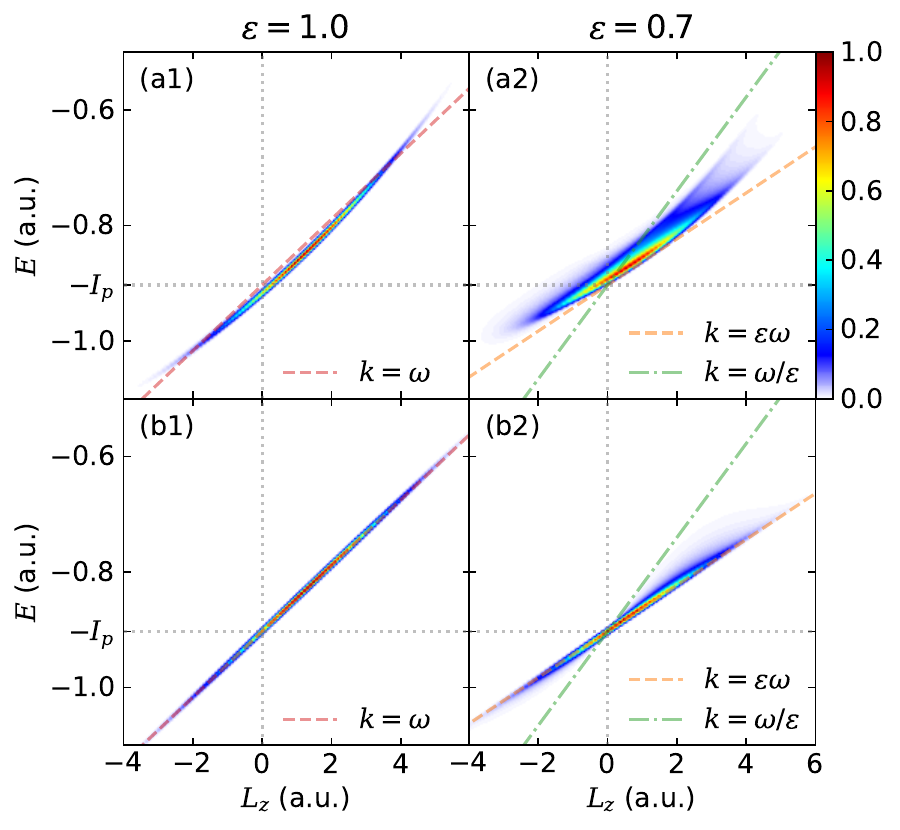}
\caption{Correlated spectrum of angular momentum and energy (SAME) at the tunnel exit for photoelectrons originating from ionization of the helium atom exposed to (1) circularly or (2) elliptically polarized laser pulses at a wavelength of 800 nm and a peak intensity of $8.0\times10^{14}$ W/cm$^2$. Row (a) presents the results obtained from backpropagation and row (b) depicts the results of SFA. The slope for the circular case is denoted by $k=\omega$ (red dashed line), and for the elliptical case, it ranges between $k=\varepsilon\omega$ (orange dashed line) and $k=\omega/\varepsilon$ (green dash-dotted line).}
\label{fig:LzEn}
\end{figure}

In the context of the present study, the differential SAME at the tunnel exit is constructed using the backpropagation method \cite{Ni2016,Ni2018,Ni2018a,Hofmann2021,Ma2023}, as shown in Fig.~\ref{fig:LzEn}(a1) for circular polarization. While the angular momentum $L_z$ ($z$ component of $\bm{L}=\bm{r}_0\times\bm{k}_\perp$ with $\bm{r}_0$ the tunneling exit position and $\bm{k}_\perp$ the initial transverse momentum, in contrast to works addressing the angular quantum number $l$ \cite{Arbo2008,Arbo2015,Popruzhenko2018}) and energy $E$ distributions exhibit broad profiles individually, their correlated distribution forms a sharp straight line, indicating a well-satisfied conservation law between the angular momentum $L_z$ and energy $E$. This linear relation closely follows the expression $E=kL_z+b$ with $k=\omega$ the slope and $b=-I_p$ the intercept, where $I_p$ is the ionization potential, leading to the conservation law for circular polarization
\begin{equation}
E=\omega L_z-I_p.
\label{eq:circle_LzEn_exit}
\end{equation}
This relation holds a transparent physical interpretation. On top of the adiabatic tunneling scenario with a constant energy $E=-I_p$, each absorbed quasiphoton results in an increase of 1 unit in the angular momentum $L_z$, accompanied by a corresponding increase of $\omega$ in the energy. It is worth noting that the computed SAME exhibits a slight downward shift and curvature with respect to Eq.~\eqref{eq:circle_LzEn_exit}, which can be attributed to the influence of the Coulomb potential of the parent ion \cite{supp}.

To gain deeper insights into the ultrafast tunneling dynamics and provide compelling evidence for the conservation law [Eq.~\eqref{eq:circle_LzEn_exit}], we carry out theoretical investigations based on the strong-field approximation (SFA) \cite{Popruzhenko2014,Amini2019}. By incorporating the saddle-point approximation \cite{Nayak2019,Jasarevic2020} within the SFA framework, we extract the tunneling exit characteristics, enabling us to construct the SAME of photoelectrons, as shown in Fig.~\ref{fig:LzEn}(b1). Notably, the relationship between angular momentum and energy precisely adheres to the conservation law [Eq.~\eqref{eq:circle_LzEn_exit}], which expression can indeed be derived fully analytically \cite{supp}.

The remarkably sharp line observed in the SAME indicates that the conservation law [Eq.~\eqref{eq:circle_LzEn_exit}] holds true at all time instances within an optical cycle, thereby supporting the notion of instantaneous quasiphotons on the subcycle scale. To further confirm this, we examine the SAME at different time points within an optical cycle, as presented in Fig.~S1 of the Supplemental Material (SM) \cite{supp}. Evidently, the SAME consistently adheres to the same conservation law throughout the entire cycle for circular pulses.

The conservation law can be naturally extended to the case of elliptical polarization. Formally, the arbitrary time translation $\hat{P}_t$ should now be accompanied by a squeezing or stretching operation $\hat{P}_s$ before applying the rotation operation $\hat{P}_\varphi$. This additional squeezing or stretching operation manifests as a varying effective angular frequency within an optical cycle \cite{supp}. To ensure a meaningful instantaneous angular frequency, such operations are limited to pulses with a sufficiently large ellipticity. In this sense, the concept of instantaneous quasiphotons is not applicable to linearly polarized light pulses \cite{supp}. Figure~\ref{fig:LzEn}(a2) displays the SAME obtained from backpropagation for an elliptical pulse with ellipticity $\varepsilon=0.7$. Not surprisingly, the SAME is no longer a sharp line but becomes broadened instead. Employing the SFA, we find that the broadening arises due to the variation in the effective energy of the quasiphoton within an optical cycle, as depicted in Fig.~\ref{fig:LzEn}(b2). Following the same line, the conservation law can now be expressed as \cite{supp}
\begin{equation}
E = \tilde{\omega}(t){L_z} - {I_p},
\label{eq:ellipse_LzEn_exit}
\end{equation}
where $\tilde{\omega}(t) = \frac{\varepsilon\omega}{\varepsilon^2\cos^2(\omega t)+\sin^2(\omega t)}$, which reduces to $\omega$ for $\varepsilon=1$. When the electric field reaches its peak at $t=T/4+nT/2$ with $n\in\mathbb{Z}$, we arrive at $\tilde{\omega}=\varepsilon\omega$; while at the electric field minima at $t=nT/2$, we have $\tilde{\omega}=\omega/\varepsilon$. As a result, the effective energy of the quasiphoton ranges between $\varepsilon\omega$ and $\omega/\varepsilon$. The subcycle variation of the quasiphoton energy is more clearly illustrated in the time-resolved plot of the SAME, as shown in Fig.~S2 of the SM. It is important to note that tunneling ionization, being a highly nonlinear process, predominantly occurs near the peak electric field. Hence, the yield maximizes for a quasiphoton with energy $\varepsilon\omega$, resulting in a time-integrated SAME concentrating primarily around $k=\varepsilon\omega$ in Figs.~\ref{fig:LzEn}(a2) and \ref{fig:LzEn}(b2).

\begin{figure}[t]
\centering
\includegraphics[width=\columnwidth]{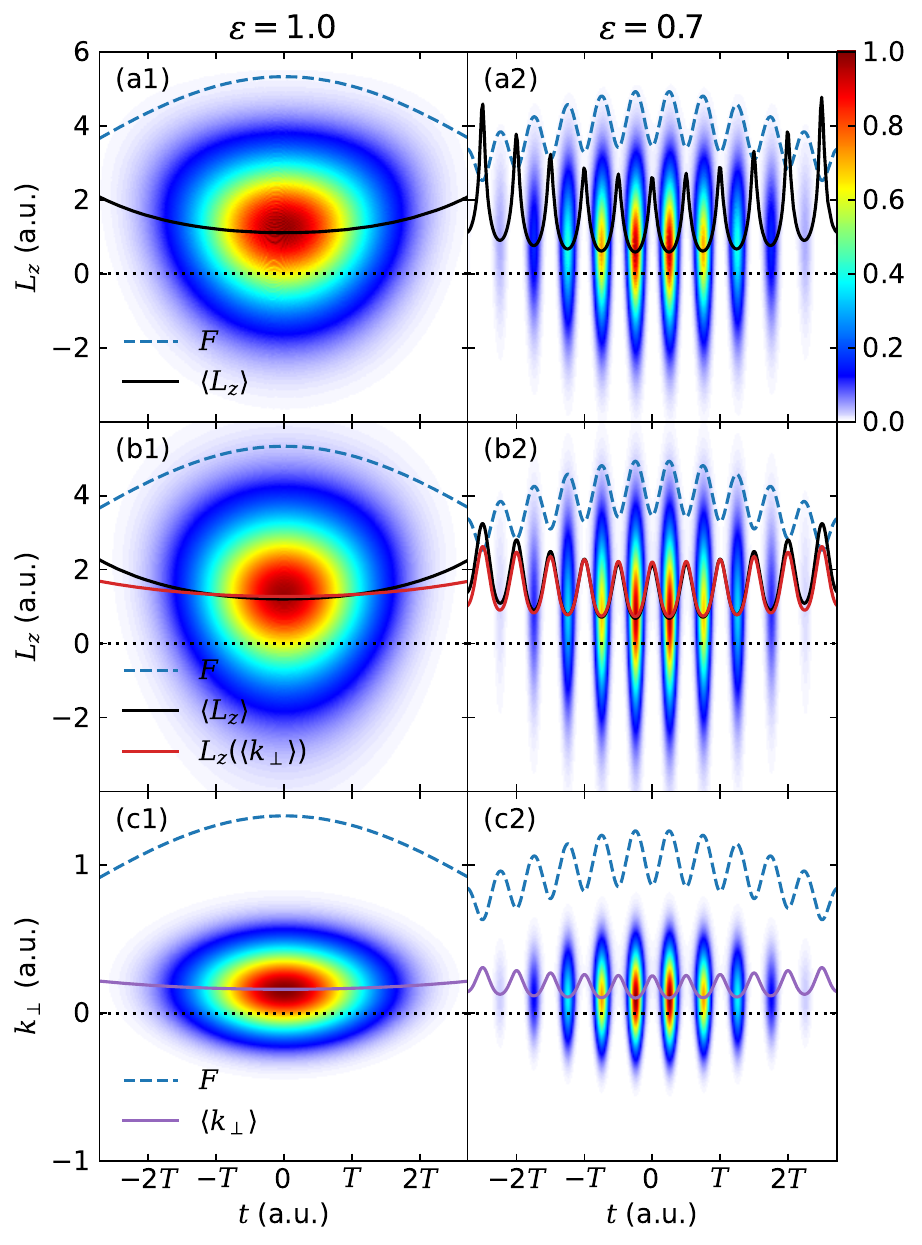}
\caption{Time-resolved distributions of the (a-b) angular momentum $L_z$ and (c) transverse momentum $k_\perp$ at the tunnel exit for photoelectrons originating from ionization of the helium atom exposed to (1) circularly or (2) elliptically polarized laser pulses. Row (a) presents the results obtained from backpropagation and rows (b-c) depict the results of SFA. The laser parameters are identical to those in Fig.~\ref{fig:LzEn}. The blue dashed line represents the magnitude of the total electric field $F(t)$, the black and purple solid lines are the time-resolved expectation value of the angular momentum $\langle L_z\rangle$ and transverse momentum $\langle k_\perp\rangle$, respectively, and the red solid line stands for the result of Eq.~(\ref{eq:Lzkd}) evaluated at $k_\perp=\langle k_\perp\rangle$.}
\label{fig:TrLzKd}
\end{figure}

\begin{figure*}[t]
	\centering \includegraphics[width=\textwidth]{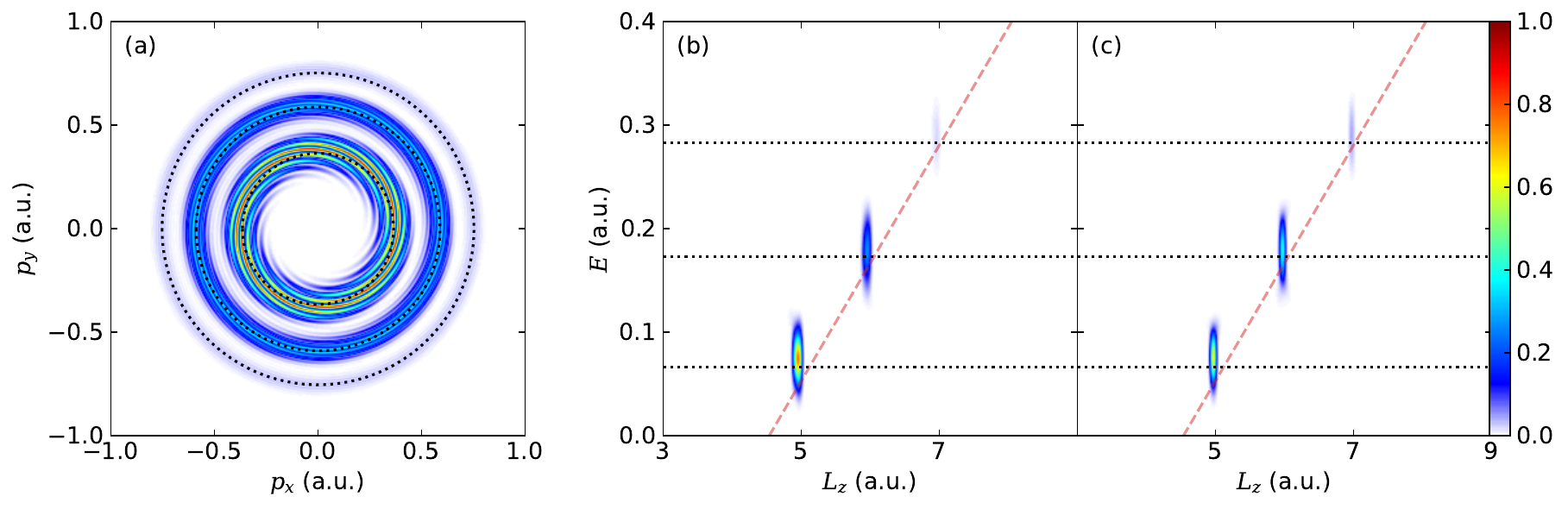}
	\caption{(a) Photoelectron momentum distribution resulting from the ionization of hydrogen atoms by the time-delayed counter-rotating circularly polarized laser pulse. (b) The SAME of the asymptotic electrons extracted from the vortex structure in panel (a). (c) The SAME of the asymptotic electrons obtained from direct evaluation of the SFA transition amplitude. The red dashed line denotes the conservation law for the asymptotic electrons $E=\omega L_z-I_p-U_p$.}
	\label{fig:LzEn_end}
\end{figure*}

Employing different theoretical approaches, we proceed to investigate the subcycle variation of angular momentum at the tunnel exit, which, to some extent, corresponds to the number of quasiphotons absorbed at specific moments. Figure~\ref{fig:TrLzKd}(a) presents the angular momentum distribution obtained from backpropagation, which evidently exhibits a broad profile. To analyze more clearly the subcycle dynamics in angular momentum transfer, we overlay on top of the distribution the time-resolved expectation value $\langle L_z\rangle$ as black solid lines. Remarkably, the average angular momentum transferred from the light to the tunneling electron oscillates inversely with the electric field magnitude, irrespective of circular or elliptical pulses. This observation is confirmed by the results of SFA in Fig.~\ref{fig:TrLzKd}(b). To isolate the origin of this behavior, we write down the expression for the angular momentum at the tunnel exit \cite{supp}:
\begin{equation}
L_z = \frac{k_\perp(k_\perp^2+2I_p)}{2\left[F(t)+\tilde{\omega}(t)k_\perp\right]},
\label{eq:Lzkd}
\end{equation}
where $F(t)$ is the magnitude of the total laser electric field. This expression reveals that at any given moment, the angular momentum depends solely on the transverse momentum $k_\perp$, whose distribution is depicted in Fig.~\ref{fig:TrLzKd}(c). Notably, the average transverse momentum $\langle k_\perp\rangle$ exhibits similar oscillations to the average angular momentum $\langle L_z\rangle$, reaching its minima when the electric field peaks. This behavior is well known to arise from nonadiabatic effects during tunneling ionization. One would thus naturally approximate the average transferred angular momentum $\langle L_z\rangle$ by evaluating Eq.~\eqref{eq:Lzkd} at $k_\perp=\langle k_\perp\rangle$, which is represented by the red solid lines in Fig.~\ref{fig:TrLzKd}(b). The overall good agreement between the black and red lines indicates that $\langle L_z\rangle\approx L_z(\langle k_\perp\rangle)$ is generally a valid approximation. We can therefore conclude that it is the nonadiabatic tunneling effects that induce the observed subcycle oscillation in the angular momentum transferred by the laser to the tunneling electrons.

Up to this point, we have focused on characterizing the properties of quasiphotons at the tunnel exit. It would be interesting if such features of quasiphotons can be accessed asymptotically as well so that it can be eventually observed. To achieve this, we need to construct the SAME of photoelectrons in the asymptotic region. While the energy can be easily obtained from the photoelectron momentum spectrum, the evaluation of the angular momentum requires phase information, which is not directly observable. In order to extract the phase distribution, an interference scheme is necessary. Specifically, we utilize two replicas of the same pulse to create a time-delayed counter-rotating circularly polarized laser pulse, which induces interference fringes in the photoelectron momentum spectrum known as electron vortices \cite{Ngoko2015,Ngoko2016,Pengel2017,Pengel2017a}. The phase information can thereby be extracted from the structure of these electron vortices \cite{Qin2021}. Further details on the phase retrieval process can be found in the SM \cite{supp}.

Electron vortices are primarily observed involving group-1 atoms \cite{Pengel2017,Pengel2017a}. These atoms have outermost orbitals of $s$ symmetry, which simplifies the interference pattern by minimizing multi-channel effects. Additionally, group-1 atoms have relatively low ionization potentials, requiring a smaller number of photons to access ATI, resulting in electron vortices with fewer spiral arms and enhanced visibility. For our purpose of phase retrieval, we perform numerical simulations using the hydrogen atom, where each individual circular pulse has a wavelength of 400 nm, a duration of 10 optical cycles, and an intensity ranging from $2.0\times10^{13}$ W/cm$^2$ to $3.4\times10^{13}$ W/cm$^2$ incorporating focal volume averaging, with a peak-to-peak time delay of 11 optical cycles. Figure~\ref{fig:LzEn_end}(a) illustrates the electron vortex generated by the time-delayed counter-rotating circularly polarized laser pulse after incorporating focal volume averaging. In the photoelectron momentum spectrum, three distinct ATI peaks (indicated by black dotted lines) are clearly visible, each characterized by a different number of spiral arms corresponding to twice the number of absorbed photons. The phase retrieval process yields a SAME shown in Fig.~\ref{fig:LzEn_end}(b), which closely matches that obtained from direct evaluation of the SFA transition amplitude in Fig.~\ref{fig:LzEn_end}(c). This demonstrates the reliability of our phase retrieval approach. In contrast to tunneling ionization, the SAME for ATI processes exhibits distinct peaks with angular momentum $L_z$ aligning sharply at discrete integer numbers corresponding to the number of photons absorbed. In contrast to that at the tunnel exit, on the other hand, the conservation law for asymptotic electrons in the circular scenario takes on a different form \cite{supp}
\begin{equation}
E=\omega L_z-I_p-U_p,
\label{eq:circle_LzEn_asymp}
\end{equation}
where $U_p$ is the ponderomotive potential. Remarkably, this conservation law resembles the equation for the ATI peaks $E_n=n\omega-I_p-U_p$, with the substitution of the angular momentum $L_z$ by the number of absorbed photons $n$. To validate the existence of instantaneous quasiphotons on the subcycle scale, we examine the phase information at various emission angles and construct the SAME for each angle. The results, depicted in Fig.~S5 of the SM, consistently adhere to the same conservation law [Eq.~\eqref{eq:circle_LzEn_asymp}] throughout the entire optical cycle, further strengthening the notion of instantaneous quasiphotons in the asymptotic region.

In summary, we have introduced in this study the concept of quasiphotons that is valid down to the instantaneous level when light is described by classical optical fields. The emergence of instantaneous quasiphotons is attributed to the infinite-order dynamical symmetry present in the interaction between atoms and circularly or elliptically polarized light pulses. We have examined the behavior of quasiphotons both at the tunnel exit and in the asymptotic region, where prominent conservation laws governing the relationship between angular momentum and energy are identified. In addition, we have proposed an approach based on electron vortices as a means for direct visualization of the existence of quasiphotons. The correlated spectrum of angular momentum and energy (SAME) associated with quasiphotons proves to be a valuable tool to assess the characteristics of photoelectrons upon strong-field ionization, especially because it encodes richer information including both the amplitude and phase of the photoelectron momentum distribution on the subcycle scale, and holds promise for further exploration and a deeper understanding of fundamental light-matter interactions.

We would like to thank Reinhard D\"orner, Sebastian Eckart, Kun Huang, and Shicheng Jiang for helpful discussions.
This work was supported by the National Natural Science Foundation of China (Grant Nos.\ 92150105, 11834004, 12227807, 12241407, 11925405, and 12274294), and the Science and Technology Commission of Shanghai Municipality (Grant No.\ 21ZR1420100). Numerical computations were in part performed on the ECNU Multifunctional Platform for Innovation (001).

\bibliography{Quasiphoton.bib}

\end{document}